\documentclass[10pt,a4paper,final]{article}
\usepackage[utf8]{inputenc}
\usepackage[english]{babel}
\usepackage{amsmath}
\usepackage{amsfonts}
\usepackage{amssymb}
\usepackage{graphicx}
\usepackage[left=2.5cm,right=2.5cm,top=2.5cm,bottom=2.5cm]{geometry}
\usepackage{color}
\usepackage{lineno}
\usepackage{hyperref}
\usepackage[style=phys,%
	biblabel=brackets,%
	chaptertitle=false,pageranges=false,%
	maxnames=4,sorting=none]{biblatex}
\usepackage{physics}
\usepackage{subcaption}
\usepackage{tikz}
\usepackage{siunitx}
\usepackage{url}
\usepackage{makeidx}
\makeindex

\graphicspath{{images/}}
\begin{document}

\begin{center}
	{\LARGE \bf Measurement of Charmonium Production in $p + p$ and $p + d$ Interactions in the Fermilab SeaQuest Experiment}

	\par\vspace*{2.5mm}\par

	{

		\bigskip

		\large \bf Ching Him Leung\footnote{E-Mail: \href{mailto:chleung@illinois.edu}{chleung@illinois.edu}\\Work supported by NSF-2111046} (on behalf of the SeaQuest Collaboration)}

	\vspace*{2.5mm}

	{Department of Physics, University of Illinois Urbana-Champaign, \\Urbana, Illinois 61801, USA}

	\vspace*{2.5mm}

	{\it Presented at DIS2022: XXIX International Workshop on Deep-Inelastic Scattering and Related Subjects, Santiago de Compostela, Spain, May 2-6 2022}

	\vspace*{2.5mm}

\end{center}
\begin{abstract}
	The Fermilab SeaQuest experiment has measured dimuon events from the interactions
	of \SI{120}{\GeV} proton beam on liquid hydrogen and deuterium targets with dimuon
	mass between \num{2} and \SI{9}{\GeV}. These dimuon events contain both the Drell-Yan
	process and the charmonium ($J/\psi$ and $\psi^\prime$) production. Unlike the Drell-Yan process
	which probes the antiquark distributions in the nucleons, the charmonium production
	is sensitive to both quark and gluon distributions. SeaQuest has extracted the
	$\sigma^{pd}/2\sigma^{pp}$ ratio as well as the differential cross sections for
	charmonium production in the kinematic region of $0.4 < x_F < 0.9$. The $\sigma^{pd}/2\sigma^{pp}$
	ratio for charmonium production are found to be significantly different
	from that of the Drell-Yan process. The measured differential cross sections for
	charmonium production are compared with theoretical calculations using Color Evaporation Model
	and Non-Relativistic QCD model.
\end{abstract}

\section{Introduction}
\label{sec:intro}
The SeaQuest experiment at Fermilab was designed to measure high-mass dimuons
produced in the interactions of \SI{120}{\GeV} proton beam with various targets.
Dimuons originating from the Drell-Yan process \cite{drell1970} as well as the
decay of quarkonium states were collected simultaneously. Result from SeaQuest
on the $\sigma^{pd}/2\sigma^{pp}$ Drell-Yan cross section ratio, which is sensitive to
the flavor asymmetry of $\bar{d}/\bar{u}$ in the proton, was reported recently
\cite{dove2021}.

Unlike the Drell-Yan process which primarily involves the annihilation of quark
and antiquark via electromagnetic interaction, charmonium production proceeds
via strong interaction containing contributions from both the quark-antiquark
annihilation and the gluon-gluon fusion processes \cite{vogt1999}.

While proton-induced charmonium production is expected to be dominated by gluon-gluon
fusion process \cite{vogt1999}, some contributions from the quark-antiquark
annihilation process is also expected. The relative importance of these two
processes is expected to depend on the energy of the colliding hadron as well
as the Feynman-$x$ ($x_F$) of the charmonium \cite{peng1995}. The quark-antiquark
annihilation process is sensitive to the light sea-quark asymmetry in the proton,
as in the case of Drell-Yan process, while the gluon-gluon fusion process is expected
to be identical for the reaction on hydrogen and deuterium targets.

The NA51 Collaboration reported the simultaneous measurement of the charmonium
production and Drell-Yan for $p+p$ and $p+d$ at \SI{450}{\GeV} at a single value
of $x_F$ \cite{abreu1998}. The SeaQuest measurement covers a broader kinematic
range of $0.3<x_F<0.8$. The \SI{120}{\GeV} beam energy in the SeaQuest experiment
is expected to probe the parton distributions at a values of $x$ different from the
NA51 experiment at \SI{450}{\GeV}.

\section{E906/SeaQuest Experiment}
\label{sec:e906}
SeaQuest is a fixed-target experiment using the \SI{120}{\GeV} proton beam
from the Fermilab Main Injector. Details of the SeaQuest spectrometer can be
found in Ref.~\cite{aidala2019}. The target system consists of seven
interchangeable targets, including a flask with liquid hydrogen, a flask with
liquid deuterium, an empty flask (vacuum), solid carbon, iron, and tungsten
targets as well as an empty space with no target (air). The targets are interchanged
periodically to reduce systematic uncertainties in the measured cross section
ratios for the different targets.

The spectrometer consists of two magnets and four tracking stations. FMag, is a
solid iron magnet that acts as a focusing magnet as well as the beam dump. It is
then followed by the first tracking station. An open air dipole magnet (KMag) is
placed between station 1 and station 2. The vertical magnetic field from both
magnets bends the muons horizontally, allowing the measurement of the momentum
of the muons. Downstream of station 3, there is a 1 m iron wall acting as a
hadron absorber. Station 4 is located behind the hadron absorber and acts as a
muon identifier. Tracks that pass through the hadron absorber and
produce hits on station 4 are identified as  muons.

\section{Extraction of \texorpdfstring{$J/\psi$}{J/psi} cross section}
\label{sec:result}
\begin{figure}[h!]
	\centering
	\includegraphics[width=0.65\linewidth]{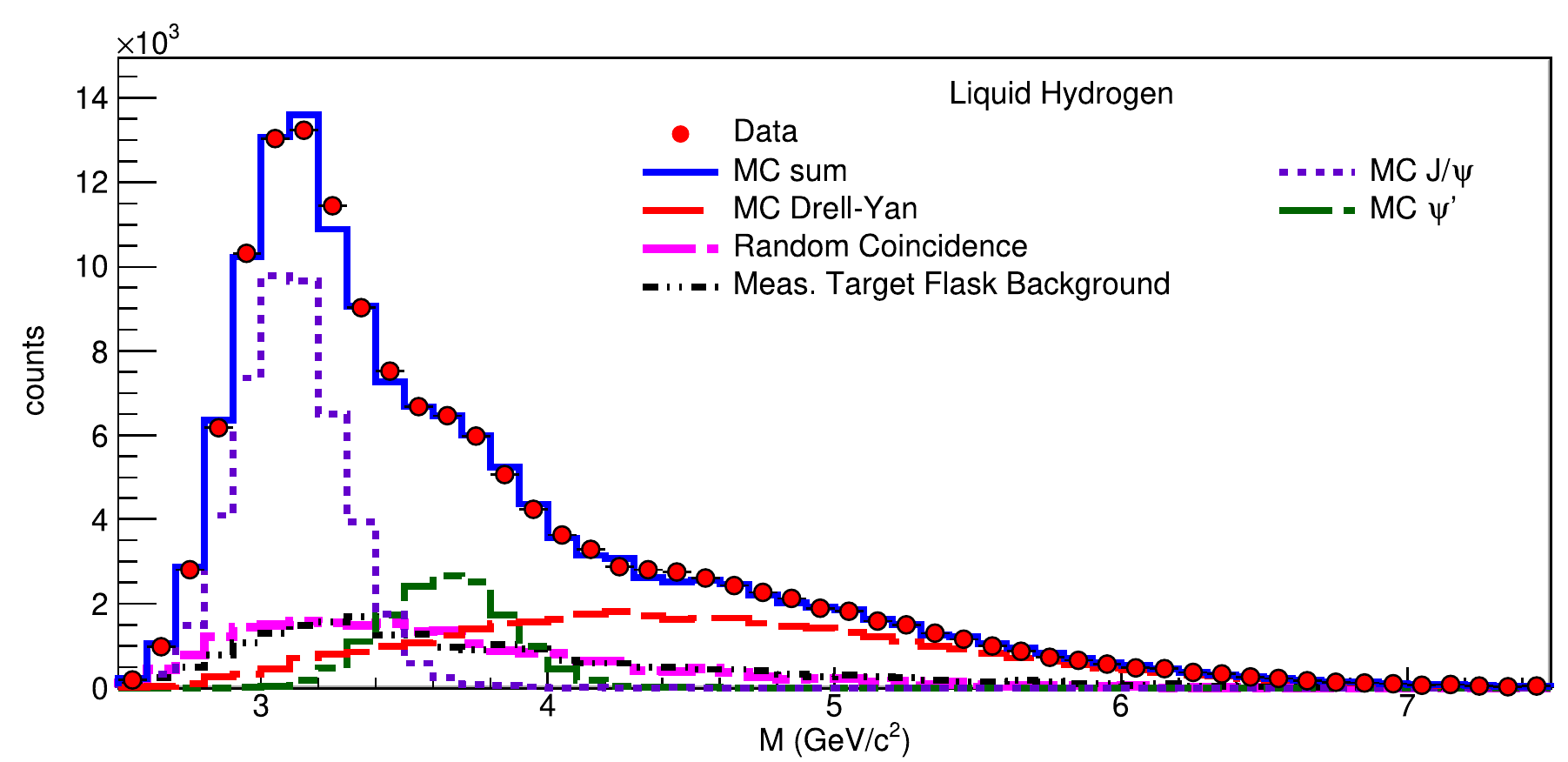}
	\caption{The reconstructed dimuon invariant mass distribution for data collected on
		liquid hydrogen target after all analysis cuts \cite{dove2021}.
		In the lower mass region, the predominant signal is from the decay
		of $J/\psi$ and $\psi^\prime$. }
	\label{fig:mass}
\end{figure}

SeaQuest experiment took data from April 2014 to July 2017. The analysis
presented here is performed on data collected until August 2015, about half
of the entire data set. After applying various analysis cuts to select
candidate dimuon events from the liquid hydrogen target, the dimuon invariant
mass distribution is shown in Fig.~\ref{fig:mass}. The $J/\psi$ peak and $\psi^\prime$
shoulder are clear visible and are the predominant sources of signal at the
lower mass region. The mass distribution is fitted with several different
components. First, the mass spectrum from the data collected with the
empty target flask, properly normalized by the integrated beam intensity,
is included. Second, the expected mass distributions for $J/\psi$ and
$\psi^\prime$, based on the analysis of the GEANT4 based Monte-Carlo
simulation are obtained.
The third component is from the analysis of the Drell-Yan
Monte-Carlo, generated using a next-to-leading order calculation and CT14 parton
distributions. Finally, the random dimuon background is simulated 
using the data collected with a ``single-muon'' trigger. Two ``single-muon''
events are combined to create a simulated ``dimuon'' events, which are sent
through the same analysis chain. The data is then fitted to a sum of these
different components. The data are well described by this fitting procedure.
This fitting procedure can also describe the mass distribution in each kinematic bin.

After the number of $J/\psi$ events is extracted from the mass distributions,
the quarkonium production cross section is obtained as follows
\begin{equation}
	B\dv{\sigma}{x_F} = \frac{N_{\textrm{events}}}{\Delta x_F \mathcal{L} \epsilon},
\end{equation}
where $B$ is the branching ratio for $J/\psi$ to decay into a muon pair,
$\epsilon$ is the spectrometer acceptance and efficiency correction, and
$\mathcal{L}$ is the effective luminosity. The spectrometer acceptance and
efficiency correction are obtained by studying the Monte-Carlo simulation.
We define the following kinematic variables for the $J/\psi$:
\begin{align}
	x_F & = \frac{2P^L_{CM}}{\sqrt{s}},                                                                                                                                                \\
	x_b & = \frac{ P_{\textrm{target}}\cdot P_{\textrm{sum}}}{ P_{\textrm{target}} \cdot\left( P_{\textrm{beam}}+ P_{\textrm{target}}\right)}, \\
	x_t & = \frac{ P_{\textrm{beam}}\cdot P_{\textrm{sum}}}{ P_{\textrm{beam}} \cdot\left( P_{\textrm{beam}}+ P_{\textrm{target}}\right)},
\end{align}
where $P^L_{CM}$ is the dimuon longitudinal momentum in the nucleon-nucleon
center-of-mass (CM) frame, $s$ is the CM energy squared. $x_b$ and $x_t$ are
the fraction of the hadron momentum carried by the parton in the beam and
target nucleon, respectively. They are defined using the four momentum of
the beam hadron ($ P_{\textrm{beam}}$), target hadron ($ P_{\textrm{target}}$)
and the dimuon ($ P_{\textrm{sum}}$).

To compare the measured $J/\psi$ production cross section with theoretical
expectations, we have performed calculation using the Next-to-Leading
order (NLO) Color Evaporation Model (CEM)\cite{mangano1993} and
Non-Relativistic QCD (NRQCD) model \cite{bodwin1997}, with CT14NLO \cite{dulat2016}
as the input PDF. In both models, the heavy
quark-antiquark ($\bar{Q}Q$) pair production via various QCD hard processes
is calculated using perturbative QCD. The main difference between the two
models is in the hadronization into specific quarkonium state.
In the CEM framework, a constant probability $F$, specific for each charmonium
state, accounts for the hadronization into colorless charmonium state. In contrast,
in NRQCD, the hadronization is described by a set of long-distance matrix elements
(LDMEs) which depend on the spin, color and angular momentum of the $\bar{Q}Q$ pairs
and the charmonium state.

\begin{figure}[h!]
	\centering
	\begin{subfigure}{0.45\linewidth}
		\includegraphics[width=0.9\linewidth]{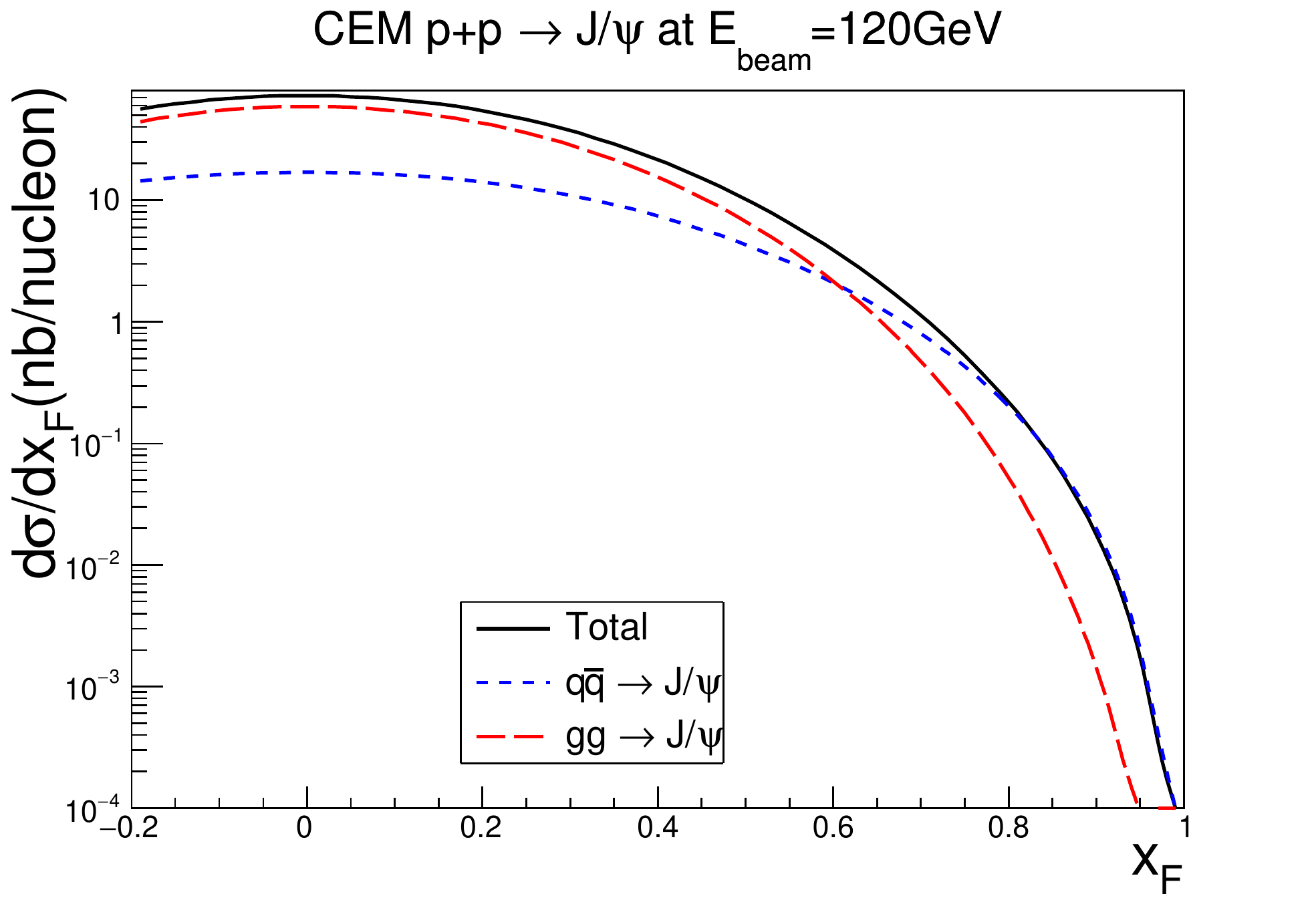}
	\end{subfigure}
	\begin{subfigure}{0.45\linewidth}
		\includegraphics[width=0.9\linewidth]{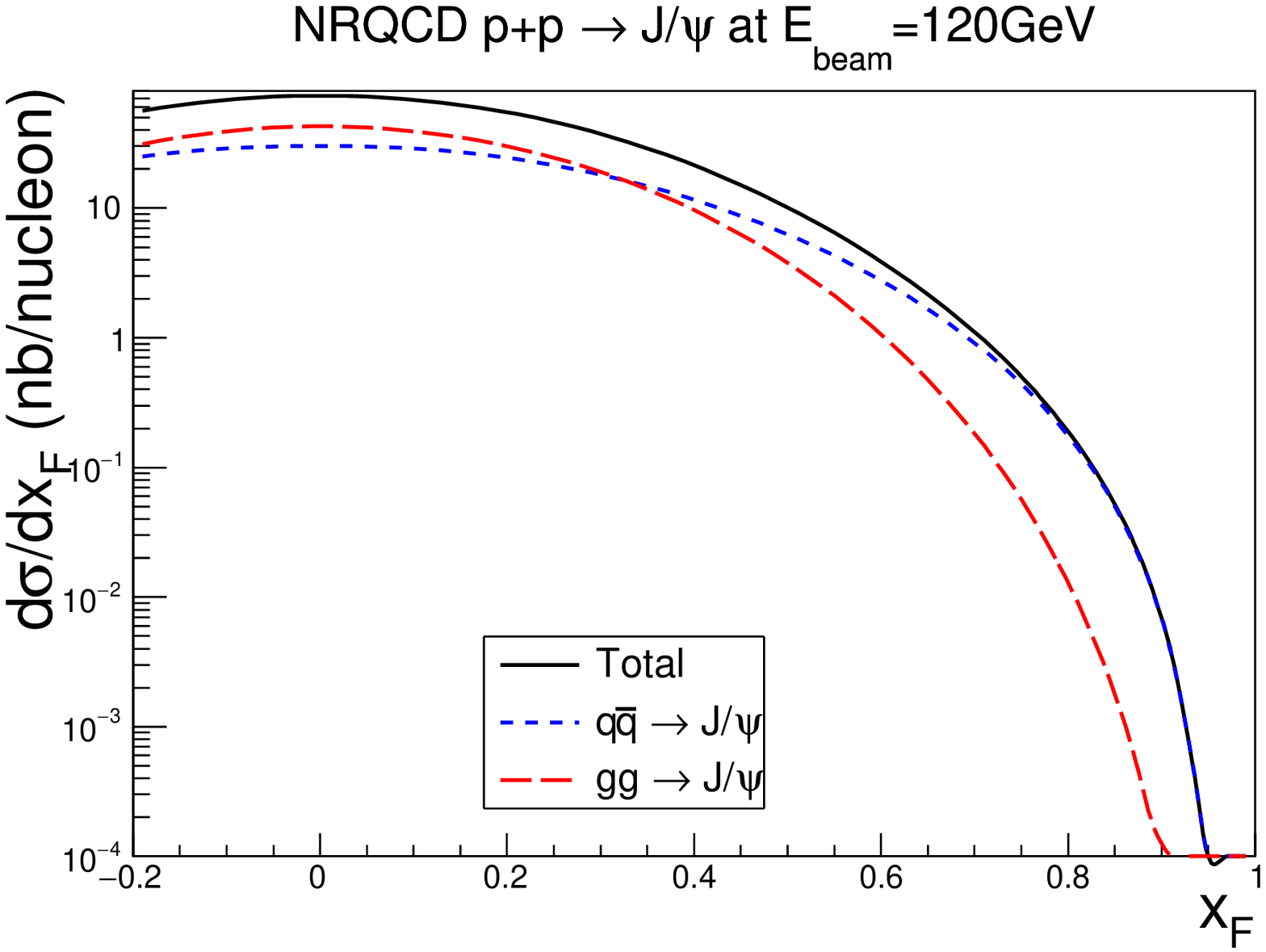}
	\end{subfigure}
	\caption{The model prediction for $J/\psi$ production in $p+p$ with a \SI{120}{\GeV}
		proton beam using CEM (left) and NRQCD (right). Contribution from the
		quark-antiquark annihilation and gluon-gluon fusion are shown as blue dotted line
		and red dashed line.}
	\label{fig:jpsi_theory}
\end{figure}

The model predictions from both models for the $J/\psi$ production in $p+p$ with a \SI{120}{\GeV}
proton beam are shown in Fig.~\ref{fig:jpsi_theory}. At forward $x_F$, both models suggest
that quark-antiquark annihilation is more important than gluon-gluon fusion, but NRQCD
gives somewhat greater importance for the quark-antiquark annihilation.

\begin{figure}[h!]
	\centering
	\begin{subfigure}{0.45\linewidth}
		\includegraphics[width=0.9\linewidth]{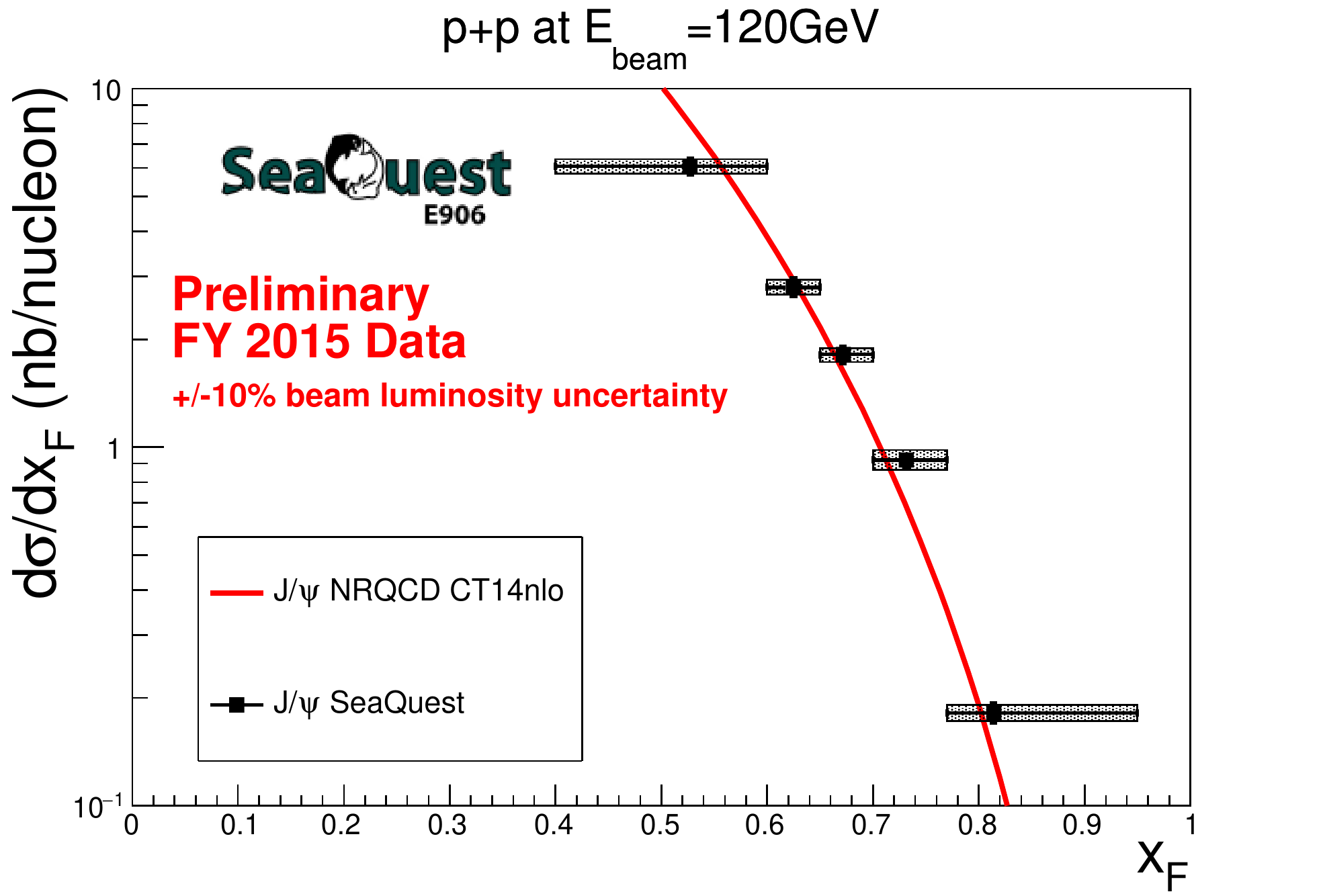}
	\end{subfigure}
	\begin{subfigure}{0.45\linewidth}
		\includegraphics[width=0.9\linewidth]{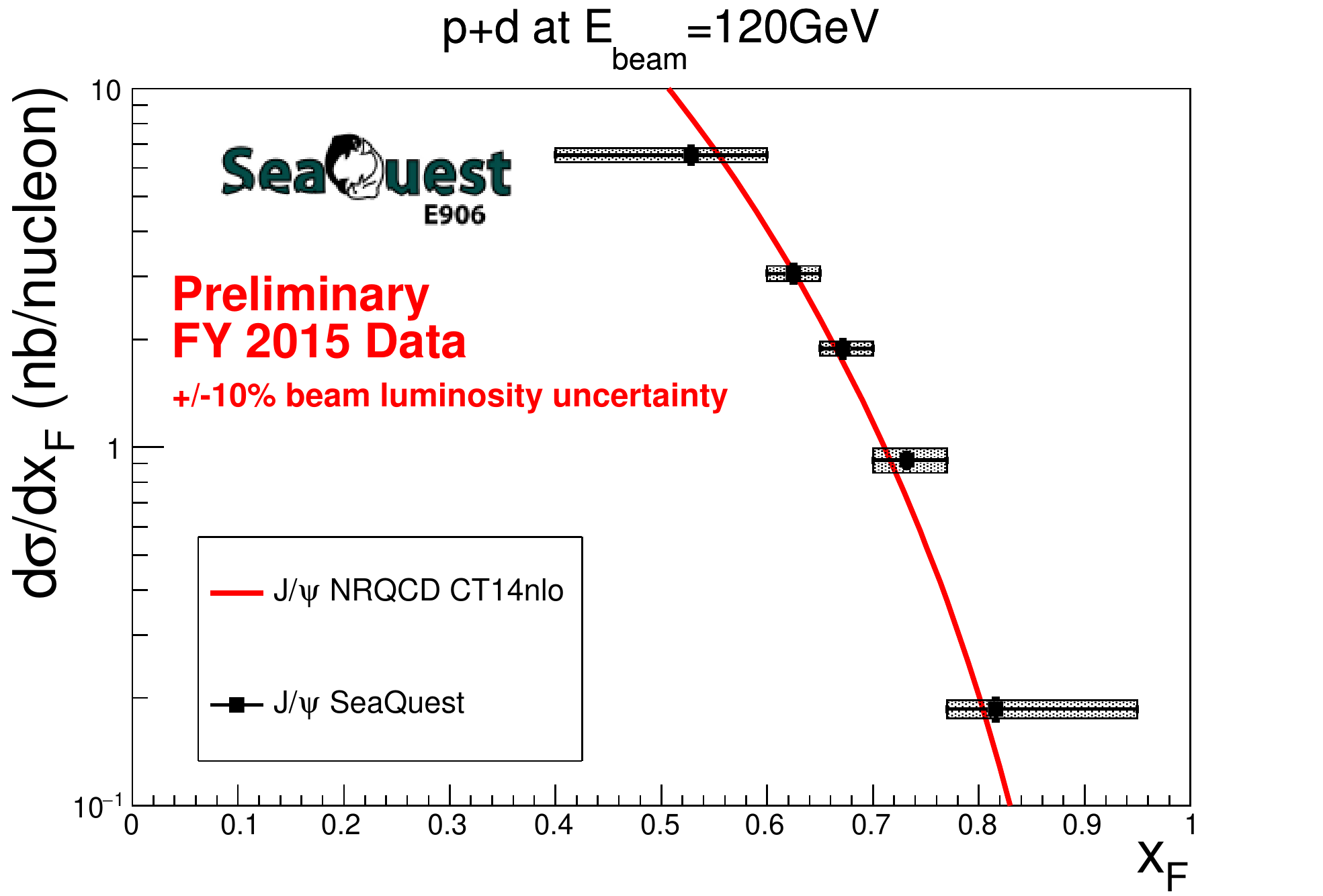}
	\end{subfigure}\\
	\begin{subfigure}{0.45\linewidth}
		\includegraphics[width=0.9\linewidth]{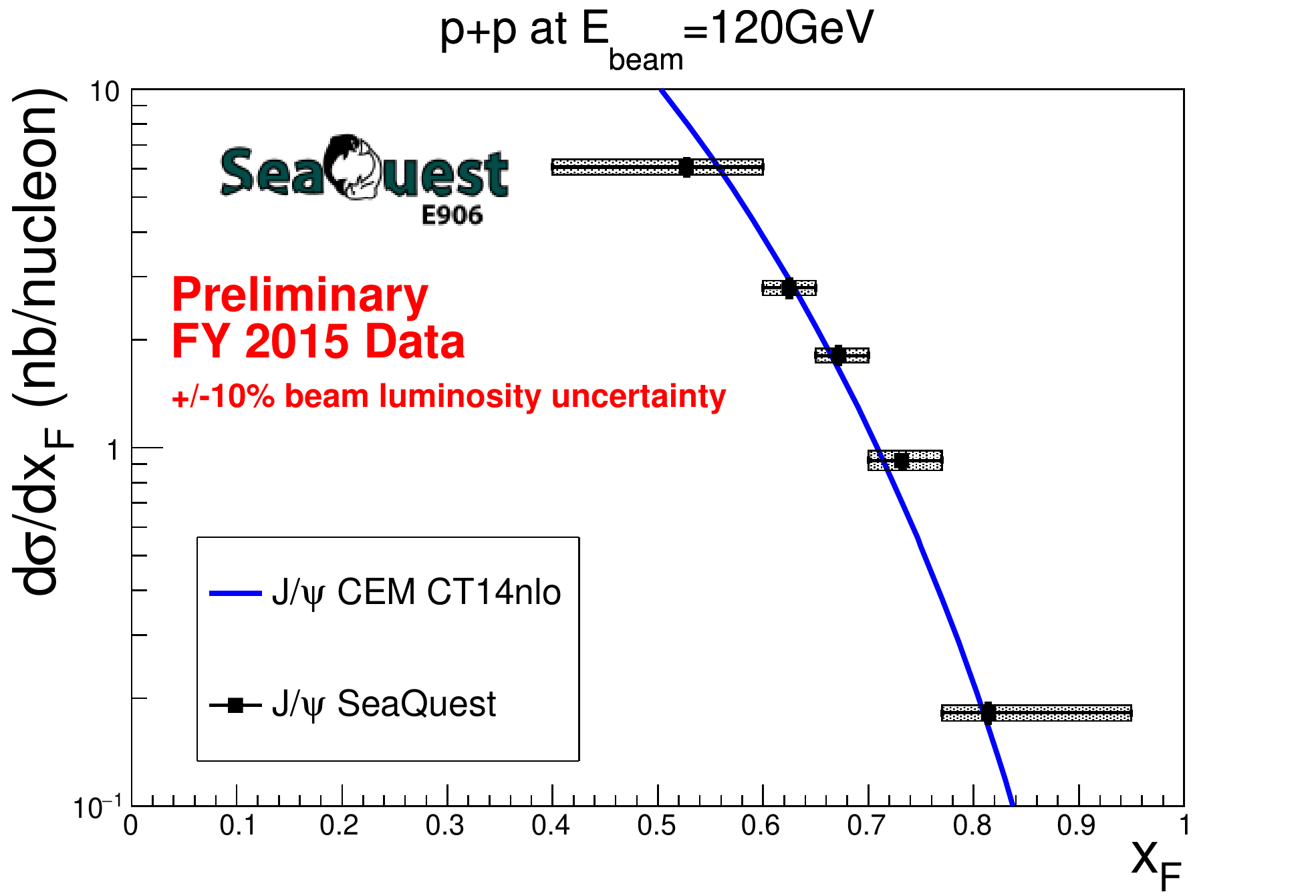}
	\end{subfigure}
	\begin{subfigure}{0.45\linewidth}
		\includegraphics[width=0.9\linewidth]{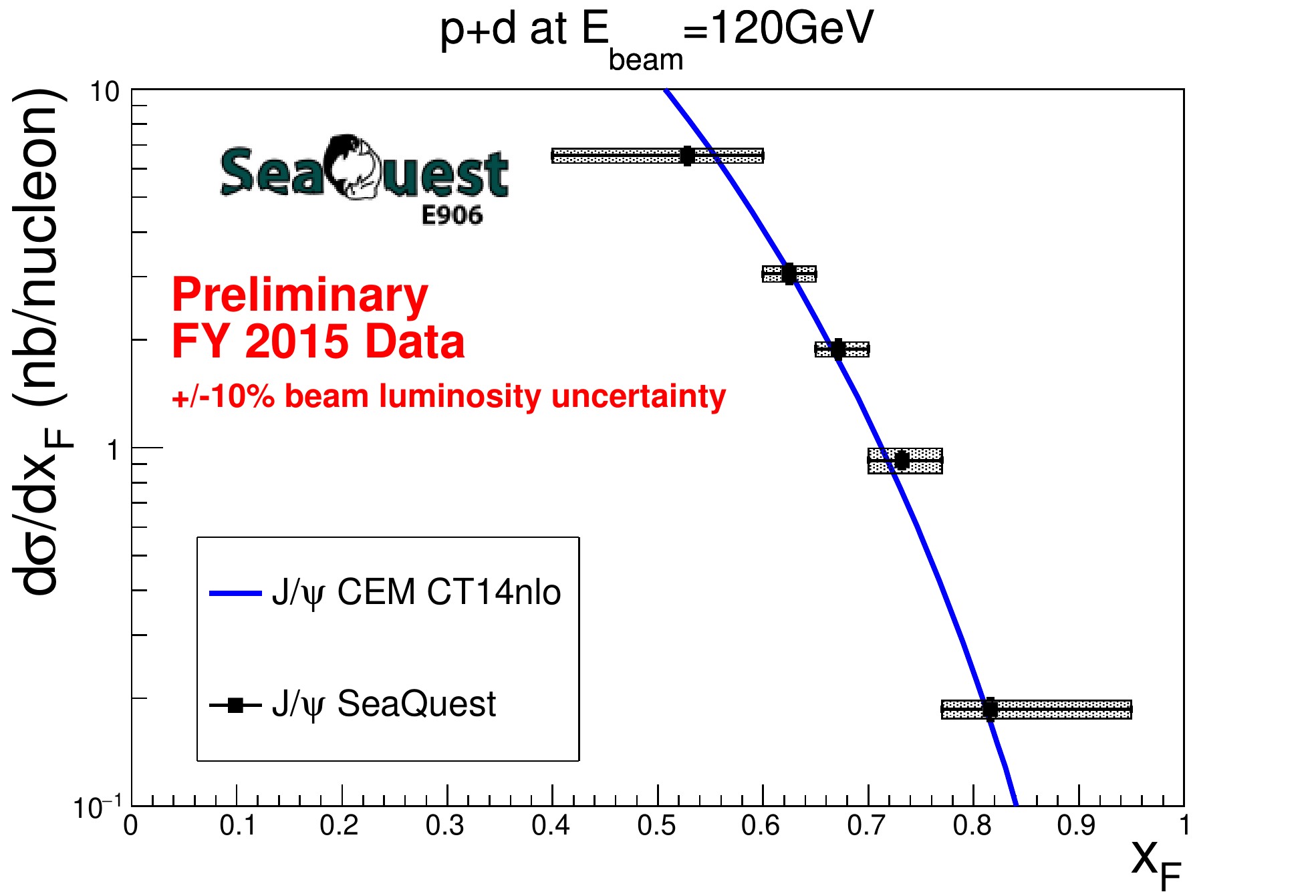}
	\end{subfigure}
	\caption{The preliminary result on the extracted $J/\psi$ cross section for
		proton on hydrogen (left) and proton on deuterium (right). The result is
		also compared with NRQCD predictions (top) and CEM prediction (bottom).}
	\label{fig:abs_cs_NRQCD}
\end{figure}

The preliminary result of $\frac{d\sigma}{dx_F}$ from SeaQuest are shown in
Fig.~\ref{fig:abs_cs_NRQCD}, and compared with both the CEM and NRQCD calculation.
The main sources of systematic uncertainties come from the modeling of the random background
and the beam luminosity normalization. The normalization of the CEM calculation,
which accounts for the hadronization probability, is adjusted to fit the data. The
shape of the extracted cross section is in good agreement with CEM. For the NRQCD
calculation, the LDMEs are taken from \cite{hsieh2021}, which are extracted from
a global analysis of pion and proton induced charmonium production in fixed-target
experiments. The magnitude and the shape of the extracted cross
sections are also in good agreement with the NRQCD calculation.

The $\sigma^{pd}/2\sigma^{pp}$ cross section ratios for the $J/\psi$ production
can also be obtained. As most of the systematic uncertainties are correlated
between the hydrogen and deuterium targets, they would cancel out in the ratio.
\begin{figure}[h!]
	\centering
	\includegraphics[width=0.6\linewidth]{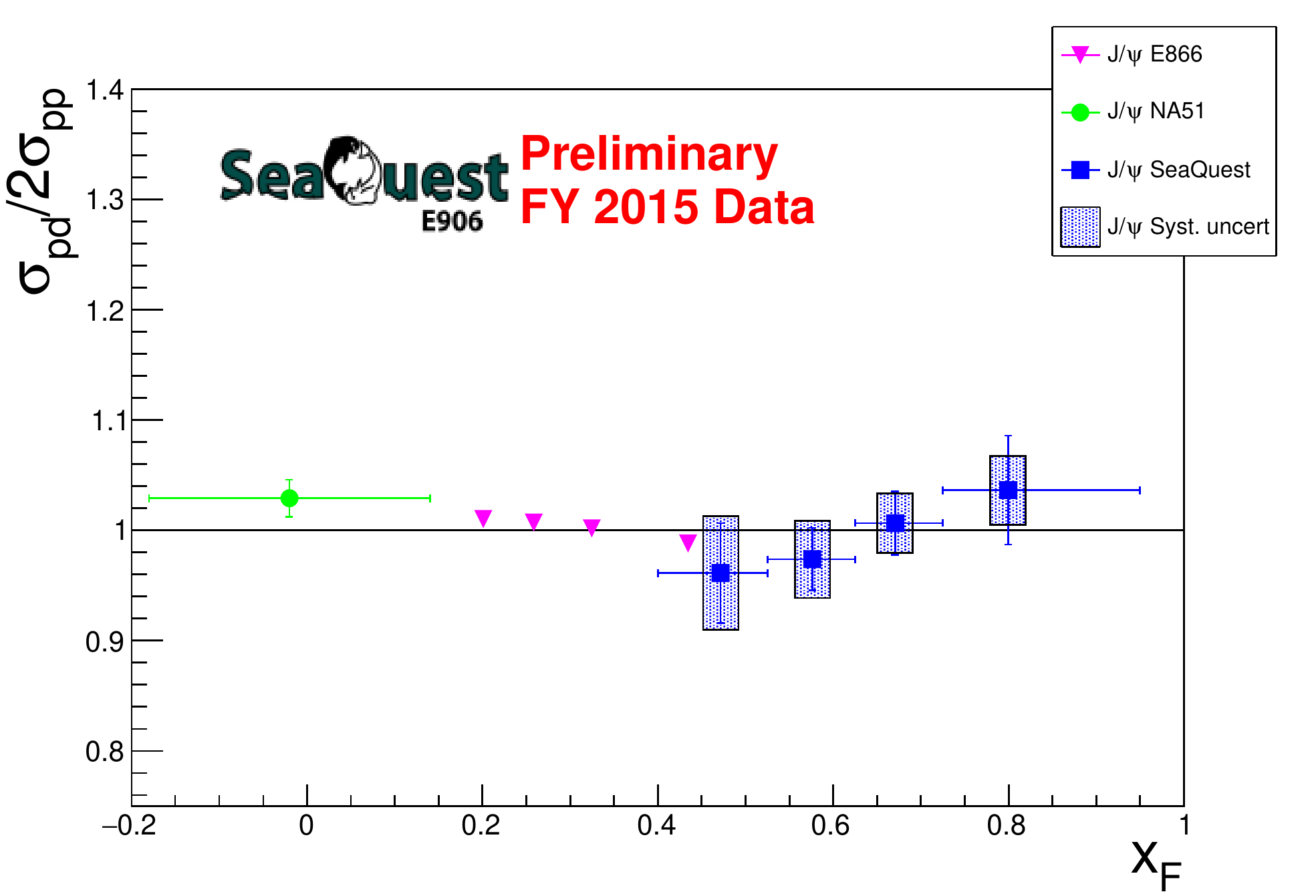}
	\caption{The preliminary result on the extracted $J/\psi$ $\sigma^{pd}/2\sigma^{pp}$
		cross section ratio as a function of $x_F$ and compared with with previous 
		measurements by NA51 and E866.}
	\label{fig:csr}
\end{figure}
\begin{figure}[h!]
	\centering
	\includegraphics[width=0.6\linewidth]{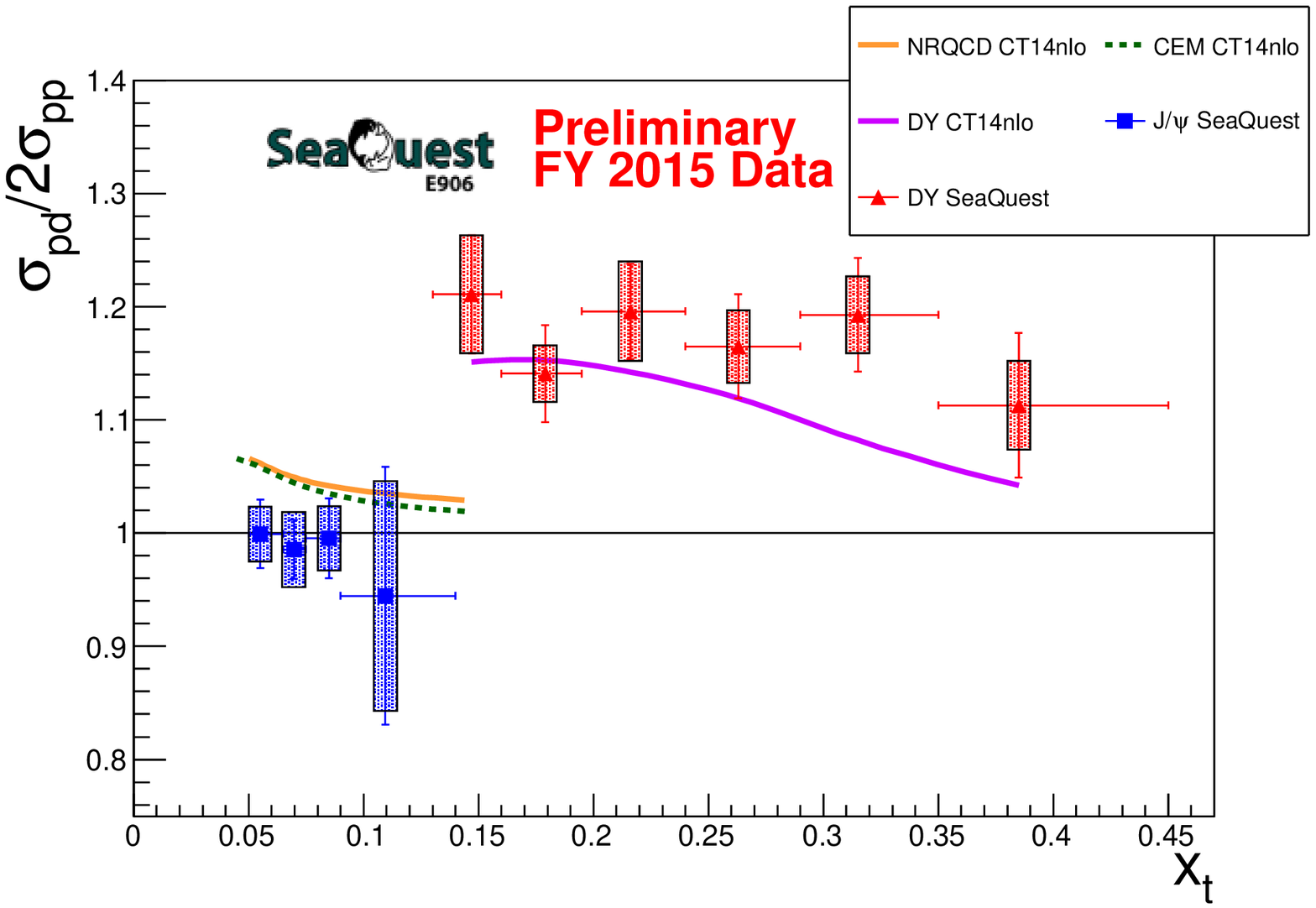}
	\caption{The preliminary result on the extracted $J/\psi$ $\sigma^{pd}/2\sigma^{pp}$
		cross section ratio as a function of $x_t$ and compared with the Drell-Yan cross
		section ratio reported by SeaQuest.}
	\label{fig:csr_xT}
\end{figure}
The preliminary $J/\psi$ $\sigma^{pd}/2\sigma^{pp}$ ratio is shown in
Fig.~\ref{fig:csr}. The SeaQuest measurement is at a lower energy and higher $x_F$
compared to previous measurements. The preliminary result is consistent with unity
within uncertainty, in qualitative agreement with earlier measurement at \SI{450}{\GeV}
by the NA51 Collaboration \cite{abreu1998} and \SI{800}{\GeV} by the E866 Collaboration
\cite{peng2003}.

Fig.~\ref{fig:csr_xT} shows the comparison of the $J/\psi$ $\sigma^{pd}/2\sigma^{pp}$ ratio
as a function of $x_t$ compared with the measured Drell-Yan ratio \cite{dove2021}.
The difference of the $\sigma^{pd}/2\sigma^{pp}$ ratios between Drell-Yan process and $J/\psi$ 
production mainly originate from the fact that the Drell-Yan process is an electromagnetic 
interaction, while the $J/\psi$ production is a strong interaction. The deviation of the cross
section ratio from unity is indicative of the light sea quark asymmetry. In the Drell-Yan
process, the sensitivity is amplified by the fact that Drell-Yan process is sensitive to the charge of the quarks.
On the contrary, the $J/\psi$ production is insensitive to the charge of the quarks, and the sensitivity to the flavor asymmetry is
further diluted by the gluon fusion process, which is expected to be identical in both $p+d$ and $p+p$.
These
different characteristics between the two processes are reflected in the
$\sigma^{pd}/2\sigma^{pp}$ ratios. While the measured Drell-Yan ratio
is significantly different from unity as a result of the $\bar{d}/\bar{u}$
flavor asymmetry of the light-quark sea, the $J/\psi$ production cross section ratio
is close to unity.
The CEM and NRQCD calculation are shown as green dotted line and orange solid line
respectively. The small difference between the two calculations reflects the greater
importance of the quark-antiquark annihilation in the NRQCD calculation as shown in
Fig.~\ref{fig:jpsi_theory}.

\section{Concluding Remarks}
The simultaneous measurement of the charmonium production and the Drell-Yan dimuons
in the SeaQuest experiment facilitates a comparison of the two distinct processes. The
extracted $J/\psi$ production cross sections are in good agreement with both the CEM
and the NRQCD calculation. The $\sigma^{pd}/2\sigma^{pp}$ $J/\psi$ production cross section ratio
is close to unity, and the difference between the $J/\psi$ and Drell-Yan
ratio is reflecting the different mechanism between the two process.

The NRQCD calculation suggest that the quark-antiquark annihilation would
be more important in the $\psi^\prime$ production, and hence $\psi^\prime$ is expected
to have a different $x_F$ distribution than $J/\psi$. The extraction of the $\psi^\prime$
cross section is currently underway and would be able to provide further input to
the understanding of charmonium production. Moreover, the data from the remaining
data sets would roughly double the statistics for both the charmonium production as well as the Drell-Yan.
The SpinQuest experiment \cite{geesaman2014}, the follow-up experiment of SeaQuest, will measure the dimuon
production with proton on transversely polarized ammonia target, which can provide information
on the transverse spin asymmetry of charmonium production.

\printbibliography[heading=bibintoc,title={References}]
\end{document}